# Spreadsheets on the Move:
# An Evaluation of Mobile Spreadsheets

Derek Flood[1], Rachel Harrison[1], Kevin Mc Daid[2]
[1]Oxford Brookes University,
[derek.flood, rachel.harrison]@brookes.ac.uk
[2]Dundalk Institute of Technology,
kevin.mcdaid@dkit.ie

**ABSTRACT**

*The power of mobile devices has increased dramatically in the last few years. These devices are becoming more sophisticated allowing users to accomplish a wide variety of tasks while on the move. The increasingly mobile nature of business has meant that more users will need access to spreadsheets while away from their desktop and laptop computers. Existing mobile applications suffer from a number of usability issues that make using spreadsheets in this way more difficult. This work represents the first evaluation of mobile spreadsheet applications. Through a pilot survey the needs and experiences of experienced spreadsheet users was examined. The range of spreadsheet apps available for the iOS platform was also evaluated in light of these users' needs.*

## 1   INTRODUCTION

Advances in technology have enabled mobile devices to allow users to accomplish a wide variety of tasks while away from traditional computing equipment (e.g. desktop or laptop computers). Traditional desktop applications are also now being ported to mobile devices, allowing users to do their computing while on the move.

One such application is the spreadsheet. With modern smart phones such as the iPhone from Apple, and the Curve from BlackBerry, users can now access their spreadsheets from anywhere. However, the small nature of these devices, required for portability has introduced a number of limitations that have caused severe usability problems. Section 2 outlines some of these issues and outlines how these issues impact upon the usability of mobile spreadsheet applications.

This paper is the first work to examine the needs of experienced spreadsheet users for accessing spreadsheets while away from traditional computing devices. By better understanding the needs of these users, more effective mobile spreadsheet applications can be developed that not only meet the needs of these users but also provide a more pleasing user experience.

A pilot survey has been conducted with experienced spreadsheet users. This survey was designed to assess not just the needs of these users but also their previous experience with mobile spreadsheets. In light of the needs of these users, existing spreadsheet applications available on the Apple iOS platform were evaluated. The research methodology employed during this study can be found in Section 3.

The results of this study have shown that there is a need for access to spreadsheets on mobile devices. Existing applications however, suffer from a number of issues that make them difficult to use and therefore provide a poor overall user experience. The full results of this study can be found in Section 4, while Section 5 outlines the lessons that have been learned during the study.

Section 6 outlines some threats to the validity of this work. The study presented here was conducted with a small number of participants which will limit the generalisability of these results. In addition to this only experienced spreadsheet users were targeted for the study. In the future a more extensive study will be conducted with all types of spreadsheet users.

In addition to this further studies will be conducted examining the usability of mobile spreadsheet applications. This study will allow for a deeper understanding of the usability issues associated with these applications. This study is outlined in Section 7 while Section 8 concludes this paper.

## 2 RELATED WORK

The rapid progression of technology has led to an increase in the number of mobile applications available. Although these applications offer a number of advantages in terms of portability and convenience they do so at the cost of usability. Zhang and Adipat (2005) have highlighted a number of issues that affect the usability of mobile applications:

- **Mobile Context**: When considering mobile applications the user is not tied to a single location. This will also include interaction with nearby people, objects and environmental elements which may distract a user's attention.

- **Connectivity**: With mobile devices connectivity is often slow and unreliable and therefore will impact the performance of mobile applications which utilise these features.

- **Small Screen Size & Different Display Resolution**: In order to provide portability mobile devices contain very limited screen size meaning that the amount of information that can be displayed is drastically reduced.

- **Limited Processing Capability and Power**: In order to provide portability, mobile devices often contain less processing capability and power. This has the effect of limiting the types of applications that are suitable for mobile devices.

- **Data Entry Methods**: The input methods available for mobile devices are difficult and require a certain level of proficiency. This problem increases the likelihood of erroneous input and decreases the rate of data entry.

The above limitations of mobile devices further aggravate existing usability issues in the spreadsheet application. The limited screen size on mobile devices requires the user to perform considerably more navigation when looking at large spreadsheets. This may cause users to find it difficult to conceptualise the overall spreadsheet and to see how the section on-screen fits with this overall picture.

Flood et al. (2008), have identified navigation as an issue that affects the performance of people debugging spreadsheets through voice recognition technology. By addressing this issue it was found that the performance of users debugging spreadsheets could be increased. It was also found that participants audited more cells with the improved navigation system, which is an important aspect of the debugging process.

Mobile devices generally do not contain a traditional keyboard as the size required would be too large to enable portability. Some devices incorporate a physical keyboard which utilises small keys while other devices use touch screen technology to present a keyboard to the user on screen. These keyboards also require the physical keys to be smaller than traditional keyboards to fit all keys on screen. The iOS platform addresses this issue by providing users with three separate keyboards; one containing letters, one containing numbers and some special characters and a third containing additional special characters.

Chen et al. (2010) conducted an evaluation of users entering text on a small size QWERTY keyboard. This evaluation required 15 participants to enter a passage of text using the small sized keyboard. On average participants used 540 keystrokes to enter the passage of text. The most prevalent type of error made by these participants during the task was a key ambiguity error, which occurred when a user entered a character other than the target character. It was found that on average, participants made about 9 key errors on the first typing task. It is also worth noting that all participants made at least one error of this type during the study.

Errors of this type, when made on a spreadsheet, may result in a misspelled word or in an incorrect reference in a cell formula, which could alter the bottom line value of a spreadsheet substantially. It has been shown repeatedly that even on desktop computers errors like this persist. Two independent studies (Panko 1998; Powell, Baker et al. 2009) have found that over 85% of the evaluated spreadsheets contained errors.

The limited processing power of portable devices has meant that existing spreadsheet applications may not function correctly when run on these devices. In an attempt to address this issue a number of developers have created spreadsheet apps which scale down the level of functionality to enable users to view and use spreadsheets in a mobile context. Most of these applications however, are limited in terms of functions available and spreadsheet size.

## 3 RESEARCH METHODOLOGY

During February 2011 a pilot survey was conducted among experienced spreadsheet users to investigate the need for mobile spreadsheet applications and to identify the issues that exist with mobile spreadsheet applications. Participants, recruited through the European Spreadsheet Risk Interest groups' mailing list, were asked to complete a short survey featuring 24 questions on their experience of using mobile spreadsheet apps and their need for such applications.

In addition to the survey, an evaluation of existing spreadsheet apps for the iOS platform was also conducted. This study examined the usability of existing spreadsheet applications as well examining the range of features they contain. The analysis of the suitability of these applications for the user's needs is presented here. A full report can be found in (Flood, Harrison et al. 2011).

### 3.1 Survey on mobile spreadsheets

The aim of this survey was to examine the extent to which mobile spreadsheet applications are required and used. The usability issues associated with existing spreadsheet applications are also investigated. To meet these aims three primary research questions were established:

- *RQ1: To what extent is access to spreadsheets required while away from traditional computing devices?*

- *RQ2: To what extent is access to spreadsheets used while away from traditional computing devices?*

- *RQ3: What issues affect the usability of mobile spreadsheet applications?*

To meet the first research question, a number of aspects were considered. In addition to asking participants if they required access to spreadsheets while away from a traditional computing devices, the survey examined the purposes for which participants required access to the spreadsheet. A deeper understanding of these purposes would allow future mobile spreadsheet applications to optimize the interface for the most common purpose, therefore improving the overall usability of the application.

A number of usability issues can be traced back to the extensive range of features found in modern applications. By identifying the most common features future spreadsheet applications could be designed so as to prioritise access to these features and make them easier to access. Although this will make other features difficult to access, it should produce a more usable application overall.

The biggest advantage of mobile applications is that they can be used anywhere, in any context. In many cases the context will impact upon the users' level of attention and it is therefore important to consider the contexts in which these applications are needed. Previous research (Schildbach and

Rukzio 2010) has demonstrated that by altering the target size of a button depending on the context of the user, the performance of the user can be increased.

While RQ1 examines the needs of experienced spreadsheet users, RQ2 examines how users have actually used mobile spreadsheet apps in the past. RQ 3 was designed to examine specific attributes of the usability of current mobile spreadsheet applications. Using participants' prior experience with mobile spreadsheet applications, the survey examined the level of satisfaction participants have had with existing mobile spreadsheet applications. Participants were also asked about what aspects of the mobile spreadsheet application they enjoyed. Including this information gave the participants an opportunity to highlight some of the positive attributes of these applications that could be utilised in developing future applications.

### 3.2 Spreadsheet App Evaluation

To better understand existing mobile spreadsheet applications, a systematic evaluation of these apps was conducted. This evaluation focused on apps available for the iOS platform available on mobile devices by Apple. It was decided to focus on this platform initially as Apple is one of the leading providers of mobile devices. It is planned to extend this evaluation to other platforms such as the Blackberry OS and Google Android. The results of this initial evaluation are summarised here in the context of participants' needs of a mobile spreadsheet application.

The evaluation protocol used during the evaluation is summarised below.

1. *Identify all potentially relevant applications.* There are a number of ways to conduct a search for appropriate applications, including a standard web search, and current software distribution methods make this increasingly easy. Most of the major mobile phone platforms now have an associated online application store. As this work is focused on spreadsheet apps for the iOS operating system, a search of the Apple App store was conducted. The search string "Spreadsheet" was used during this search.

2. *Remove light or old versions of each application.* Many software developers release trial versions of their systems, which are often free. Some of these versions include only a subset of the functionality offered by the full application whilst others allow full access to the application but for a limited time period. These types of applications should be removed if the full version of the app is also included within the search results.

3. *Identify the primary operating functions and exclude all applications that do not offer the required functionality.* The primary operating functions include frequently used functions and also occasionally used functions that are essential for the correct operation of the system in a desired context. For example, the initial system setup might include language and currency settings that would depend upon the country of use. The primary functionality of interest is to allow a user to perform spreadsheet tasks on a mobile device.

4. *Identify all secondary functionality within the remaining apps.* In addition to the primary operating functions, mobile apps will offer users a range of secondary functionalities which can enhance the application. A thorough knowledge of these functions will enable the application developers to see what functionality is available and may present opportunities for additional functionality to be included in future applications.

5. *Install each of the remaining applications, and test each of the tasks using Keystroke level modelling.* Keystroke Level Modelling (KLM) is a well established technique (Card, Thomas et al. 1983) for estimating the time taken to complete certain tasks. This will provide a quantitative measure of the difference between the efficiency of applications. KLM was used to measure the average number of interactions required to enter each of a set of 5 single digit numbers as well as the number of interactions to enter a subsequent formula to total these five numbers.

# 4 RESULTS

## 4.1 Survey Participants

In total fourteen participants took part in the online survey. The participants were recruited through the European Spreadsheet Risk interest groups' mailing list which is comprised of approximately 750 email addresses, featuring members from both academia and industry, who share in interest in the use of spreadsheets and the risks associated therein.

All participants use spreadsheets on a daily basis. When asked which of the following options best describes their level of expertise with spreadsheets; Novice, Intermediate or Expert. 71% of participants rated themselves as experts while 21% rated themselves as Intermediate. The remaining 7% rated themselves as novice spreadsheet users.

To better understand the nature of the participants' use of spreadsheet participants were asked "*In the last week, approximately how many times did you use a spreadsheet*". The responses to this question varied widely, from 6 to over 100 times. As the amount of time spent using a spreadsheet can vary per usage, participants were also asked to estimate how long they spend on average per use of a spreadsheet. Participants were asked to select one of the following options or to enter their own value: *Less than 30 minutes, 30 – 60 minutes, 1 – 2 hours, 2 – 3 hours, 3 – 4 hours, over 4 hours*. It can be seen from Figure 1 that most participants spend between 30 – 60 minutes per spreadsheet use.

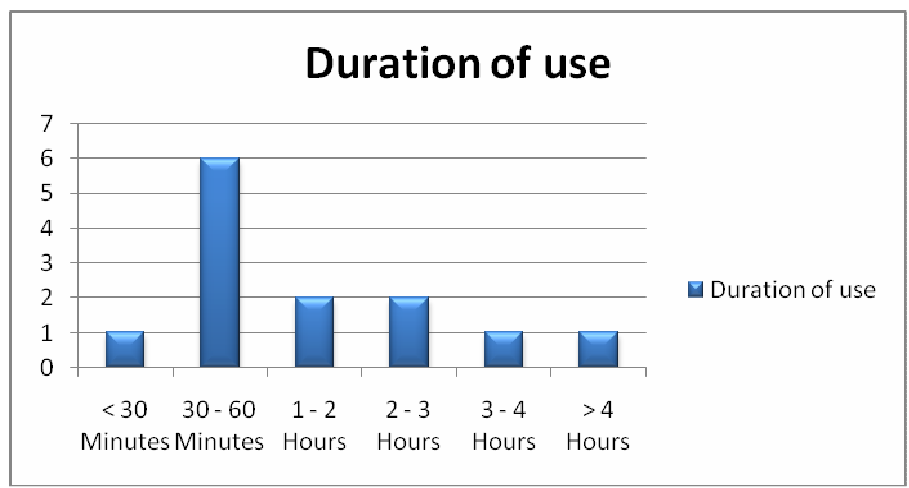

**Figure 1: Average Duration of spreadsheet use**

## 4.2 Survey Results

**The need for mobile spreadsheets**

*RQ1: To what extent is access to spreadsheets required while away from traditional computing devices?*

Approximately 79% of participants said that they required access to a spreadsheet while away from a desktop or laptop computer. These results indicate that there is a strong need for mobile spreadsheet applications.

*Purpose of spreadsheet app usage*

It is important to consider the reasons why mobile spreadsheets are to be used. To identify this, participants were asked "*for what purpose did you need it* [access to a mobile spreadsheet while away from traditional computing devices]". Participants were asked to select all that applied from the following options: *To view spreadsheet data*, *To change a spreadsheet* and *To create a new spreadsheet*.

Table 1 shows the percentage of participants who required a mobile spreadsheet application for each purpose. It can be seen that the majority of participants required spreadsheets for viewing and/or changing an existing spreadsheet. Only 21.4% of respondents said that they needed to create a spreadsheet on a mobile device.

| Purpose | % of respondents |
|---|---|
| To view spreadsheet data | 64.3 |
| To change a spreadsheet | 64.3 |
| To create a new spreadsheet | 21.4 |

Table 1: Percentage of participants by spreadsheet purpose

A critical component of spreadsheet applications is the ability to transfer data to and from the device easily. Existing spreadsheet apps for the iOS platform allow users to transfer data in a number of ways. The easiest way for users to transfer spreadsheets to the mobile device is through email. When participants receive an email including a spreadsheet, they can open the spreadsheet directly from within the email application. In addition to this some spreadsheet apps allow users to email spreadsheets directly from within the app.

Other methods of data transfer require users to be connected to the device containing the spreadsheet. The most common means is through Wi-Fi, where users can access the mobile device through a web based interface which allows users to transfer files between devices. This approach requires users to type in a specific IP address into the browser after ensuring both devices are connected to the same wireless network.

One final approach offered by a small number of apps (25%) is the use of online storage, such as Google spreadsheets. These applications allow users to log into their Google account and download their spreadsheets to their mobile device. This approach supposes that users have such accounts and store their spreadsheets in these locations. One of the applications evaluated does not allow users to transfer data to and from the device and is solely designed for use on the device.

A number of applications provide access to on-line spreadsheet solutions such as Google spreadsheets. As the access to the internet is not always available on mobile devices these applications were not considered. All of the applications evaluated allow users to operate the spreadsheet on the device alone once it has been downloaded.

*Context of spreadsheet usage*

The biggest advantage of mobile devices is the range of contexts in which they can be used. Whether out walking or travelling to work, users can access a wide variety of services that allow them to accomplish a range of tasks. In order to determine in what context mobile spreadsheets would be used participants were asked "*In what context did you need access to the spreadsheet where no desktop or laptop computer was available*". Participants provided a wide variety of contexts in which they needed access to spreadsheets where traditional computers would be impractical. Some of the contexts cited are presented below:

> *"Discussing changes in assumptions"*

> *"Could not get to laptop - but had iphone with me."*

> *"Inbound email onto a mobile device"*

> *"Daily commute"*

> *"To demonstrate data to clients"*

The broad range of contexts outlined above is from a small population sample. It is believed that a more extensive survey will produce an even broader range of contexts in which spreadsheets are required. The context in which the application is being used is an important factor as this will have an impact upon the appropriate design of the system.

*Required features on a mobile spreadsheet app*

As mobile devices are limited in terms of processing power, a full spreadsheet application would be slow and resource intensive. Most mobile spreadsheet applications therefore only supply a subset of the available spreadsheet functionality. In order to determine which functionality is important to them, participants were asked which features they required.

| Feature | % of participants |
|---|---|
| Formula | 64.3 |
| Data Features | 57.1 |
| Graphs/Charts | 28.6 |

**Table 2: Features Required**

Table 2 shows the percentage of participants who needed each feature. It can be seen the most needed feature is formula. Functions can be used to simplify the creation of such formulae. A broad range of functions are available within the mobile spreadsheet apps evaluated. The number of functions varies considerably by app, with some offering a single function, while others offer up to 146 functions. The range of financial functions available is quite limited, with most applications only offering less than 5 functions. *MarnerCalc* is an exception to this offering users access to 18 financial functions.

Only 25% of the apps evaluated allow users to use graphs and charts. Other features such as the sort function are more common among the apps, appearing in 50% of those evaluated. These results would indicate that existing applications are focused on real users' needs. It should be noted however that none of the apps evaluated featured the ability to filter data.

In addition to the features included in Table 2, some participants also stated that they required Visual Basic for Applications (VBA) macros, which are often used to enhance the functionality of spreadsheets. During this evaluation of mobile spreadsheet apps however, it was found that none of the applications offered this functionality.

*Frequency of mobile spreadsheet apps*

| Frequency | % of participants |
|---|---|
| Daily | 0.0% |
| Weekly | 21.4% |
| Monthly | 14.3% |
| Less than once a month | 50.0% |
| Never | 14.3% |

**Table 3: Frequency by which access to mobile spreadsheets are needed**

Table 3 shows the frequency with which participants have needed access to spreadsheets while away from a desktop or laptop computer. It can be seen that most participants needed access to the spreadsheet less than once a month. 21.43% of the participants did say that they needed this access on a weekly basis. These findings would indicate that memorability is an important aspect of the usability of mobile spreadsheet apps, as most participants use them on an infrequent basis.

**The use of mobile spreadsheets**

> *RQ2: To what extent is access to spreadsheets used while away from traditional computing devices?*

In addition to examining the participants' need to access spreadsheets while away from desktop or laptop computers, the survey also examined participants' previous experience using spreadsheets on a mobile device. Of the 14 participants included within the study only 42.9% said they have previously accessed spreadsheets on a mobile device. The following results are based on only these participants.

*Purpose of spreadsheet app usage*

| Purpose | % of participants |
|---|---|
| To view spreadsheet data | 83.3% |
| To change a spreadsheet | 50.0% |
| To create a new spreadsheet | 0.0% |

**Table 4: Purpose of mobile spreadsheet usage**

Table 4 shows the percentage of participants who have used a mobile device for each purpose. These results indicate that the most common use of spreadsheets on a mobile device is for viewing existing spreadsheet data. Only half of the participants have used spreadsheets for changing a spreadsheet. None of the participants have created a spreadsheet on a mobile device. These results highlight the need for mobile spreadsheet apps to allow users to transfer and access existing spreadsheets easily.

*Context of spreadsheet usage*

As was seen when examining the need of mobile spreadsheet applications, participants have used mobile spreadsheet applications in a broad range of contexts for a wide variety of tasks. One participant remarked that they have used a mobile device to change assumptions to perform sensitivity analysis while another participant wanted to read a spreadsheet that had been received in an email. Participants have also used mobile spreadsheet applications on their daily commute, a task that may be impractical with conventional laptop computers, due to their cumbersome nature in certain environments (public transport for example).

*Used features on a mobile spreadsheet app*

| Feature | % of participants |
|---|---|
| Formula | 66.7% |
| Data Features | 50.0% |
| Graphs / Charts | 16.7% |

**Table 5: Features Used**

The features used by participants on a mobile spreadsheet app are presented in Table 5. It can be seen that the most common feature used by the participants is formulae, a central component of a spreadsheet. Data features such as sort and filter options were used by half of the participants. Only 16% of participants used graphs or charts on a mobile device. These results would indicate that more complex features such as graphs and charts are less important to participants than more basic functionality such as formula. The use of data features is typically associated with only certain spreadsheets, i.e. those which contain tables of data. The relatively high frequency of their use would indicate that participants are using this type of spreadsheet frequently on mobile devices.

*Frequency of use*

When asked how often they use spreadsheets on a mobile device, 33% of participants who have used spreadsheets on a mobile device said that they use them on a weekly basis. The majority of these participants however replied that they would only need spreadsheets on a mobile device less than once a month. This infrequency of usage would indicate that the use of a mobile spreadsheet app should be memorable. There are a number of ways in which this could be achieved, for example by replicating existing interface elements from more traditional spreadsheet applications.

| Frequency | % of participants |
|---|---|
| Daily | 0.00% |
| Weekly | 33.33% |
| Monthly | 0.00% |
| Less than once a month | 66.66% |
| Never | 0.00% |

Table 6: Frequency of mobile spreadsheet usage

**The usability of mobile spreadsheets**

*RQ3: What issues affect the usability of mobile spreadsheet applications?*

A number of definitions of usability exist. The ISO define usability as *"Extent to which a product can be used by specified users to achieve specified goals with effectiveness, efficiency and satisfaction in a specified context of use"* (ISO/IEC 1998). However, Nielsen (Nielsen 1993) also identified a number of attributes that this definition does not include but are important in terms of usability; Learnability, Memorability and Errors.

The authors have identified seven attributes of usability of mobile applications; Effectiveness, Efficiency, Satisfaction, Learnability, Memorability, Simplicity and Cognitive load. Due to the nature of the research method it was not possible to evaluate participants' experience of all of these attributes. However, the following attributes of usability were examined; Effectiveness, Efficiency, Satisfaction, and Simplicity.

*Effectiveness*

To evaluate *effectiveness* participants were asked if they could accomplish their desired task using the mobile spreadsheet application. Participants were asked to select one of three options to answer this question; *Yes*, *No* or *Partially*.

| Could complete | % of participants |
|---|---|
| Yes | 33.33% |
| No | 16.67% |
| Partially | 50.00% |

Table 7: Effectiveness of mobile spreadsheet apps

Table 7 shows that only 33% of the participants could accomplish their desired task using a mobile spreadsheet application. In addition to this a further 50% could partially complete their desired task. Only 16% of the participants were unable to complete their task. These results indicate that the standard of mobile spreadsheet applications is low for the tasks that participants wish to accomplish. If mobile spreadsheet apps are to be successful their effectiveness will need to be increased dramatically.

*Efficiency*

The efficiency of participants was not evaluated through the online survey. Participants did describe using spreadsheets on a mobile device as being tedious; indicating the efficiency of these applications is quite poor. During the evaluation of existing spreadsheet apps, the efficiency of these applications was measured using a KLM analysis (Card, Thomas et al. 1983) of creating a simple spreadsheet.

This evaluation found that the efficiency of these applications varied by as much as 100%. To input 5 single digit numbers and then total these numbers in the evaluated apps took between 19 and 36 keystrokes depending on the application. To enter a single digit on some applications took up to 4 keystrokes. The reason for this is that these applications require users to select a cell, open the keyboard, select the numerical keypad and then enter the number. When the application then moves to the next cell the keyboard is automatically closed forcing users to reopen the keyboard when on this cell.

It was found that some of the spreadsheet apps evaluated allowed users to use shortcuts to enter the sum formula. When selected from a shortcut menu item, the system will automatically infer that the user wants to sum all the numbers within the selected column and will enter the formula to do so. Other applications however require the user to manually type in the formula, which can be tedious and time consuming on the touch screen.

During the evaluation it was found that one spreadsheet app had reversed the usual presentation of numbers and letters to denote the rows and columns. While most traditional spreadsheet applications number the rows and identify columns through the use of letters, this app used letters to denote the rows and numbered the columns. By altering the users' preconceptions, a new potential of error is introduced especially when users are switching between conventional spreadsheet applications and mobile spreadsheet apps.

*Satisfaction*

The satisfaction of users was measured on a five point Likert scale, where 1 indicated very satisfied and 5 indicated very unsatisfied. Of the six participants who had used mobile spreadsheet apps, the average rating was 3.33, indicating that participants were slightly dissatisfied with the existing applications. There are a number of reasons why users are dissatisfied with these applications.

When asked if they experienced any errors while using the mobile spreadsheet app, 50% of the participants said that they did experience errors. The remaining participants answered no to this question. However a subsequent question asked participants to list all of the problems that they experienced. 83% of the participants listed at least one problem for this.

The most common problem quoted by the participants related to the size of the device. Most participants had problems viewing large spreadsheets on the small screen. One participant also found it difficult to progress with their task. Another participant described spreadsheets on a mobile device as being *Useful but tedious.*

Despite the issues outlined above, some participants had some positive comments to make about the use of spreadsheets on a mobile device. The most common positive comment related to the ability to view the spreadsheet on the mobile device. The high frequency of this type of comment would indicate that participants had low expectations of spreadsheets on these devices and were pleased that they at least existed and therefore are willing to put up with a lot of errors and limitations without it having a large negative impact on the satisfaction they had with the device.

*Simplicity*

Simplicity refers to how well a user can complete their task without errors. When asked if they made any errors while using a mobile spreadsheet app, 50% of participants confirmed that they did. In addition to this participants were also asked if the device was sufficient for their needs. Half of the participants said that the device was sufficient while half disagreed saying it was insufficient. The main reason participants found it insufficient was that the size of the screen was too small. One participant also remarked that there was no access to VBA and therefore the device was insufficient to meet their particular needs.

## 5 SUMMARY

It has been found that there is a strong need for mobile spreadsheet apps. Approximately 79% of the participants surveyed said that they have needed access to a spreadsheet while away from a traditional computing device. The rapid increase in the power of mobile devices and the relatively low development costs associated with mobile applications has enabled the spreadsheet application to be ported to mobile devices.

The limitations imposed by the miniature nature of these devices has meant that mobile spreadsheet applications suffer from a number of usability issues which limit the usefulness of these applications and make them tedious for users to use. The biggest issue identified by participants has been the very small screen size of these devices, meaning that users can only view a small portion of a spreadsheet at a time. The severity of this issue could be reduced through the use of a *mini map*, which shows a scaled down version of the spreadsheet with the current area being displayed highlighted in the bottom right hand corner of the screen. Similar approaches have been used for looking at images, web pages and other large documents (Burigat, Chittaro et al. 2008).

It was also found that despite the many limitations of mobile spreadsheet applications, users were only slightly dissatisfied with the existing mobile spreadsheet applications. This result shows that the convenience of accessing the spreadsheet on a mobile device out-weighs the negative usability aspects experienced.

Despite the limitations outlined above, spreadsheets can be used on mobile devices. It was found that the most efficient app was Spreadsheet by AppAuthours [http://www.appauthors.com]. This application allowed the test spreadsheet to be created with the fewest interactions, and offers users the ability to sort data and to use freeze panes to keep headings displayed when looking at data. This app however does not allow users to view or create charts.

## 6 THREATS TO VALIDITY

The results presented above are based on a small number of participants which limits the generalisability of these results. It is hoped to conduct a more extensive survey in the future, sampling not just experienced spreadsheet users but also novice spreadsheet users. This broader sample will enable a more general picture of the need for mobile spreadsheet apps to be attained.

The pilot survey outlined above contained a limited subset of the existing spreadsheet features. A more detailed set of features should be used to determine which features are required by users and which features would not be beneficial if included within a mobile spreadsheet app. Similarly an investigation into the functions most needed by participants would enable mobile apps to better facilitate users' needs.

The subjective nature of surveys will limit the accuracy of the reported frequency of errors made by participants on mobile devices. A more accurate measure would be to monitor the participants while using a mobile spreadsheet app and record the errors that are made by participants. This will allow for a more detailed understanding of the errors that are made while using mobile spreadsheet apps.

The evaluation of existing mobile spreadsheet apps is limited to those available on the iOS platform. Other mobile platforms such as the RIM featured on Blackberry devices, Android from Google and Windows Mobile may also feature additional spreadsheet applications which would need to be evaluated to provide a complete picture of mobile spreadsheet applications. The devices on which these platforms operate will also vary, providing different modes of input; such as traditional keyboard or touch screen or a combination of both.

## 7 FUTURE WORK

The pilot survey presented above has shown that there is a need for mobile spreadsheet apps. It is intended to run a more extensive survey including not just experienced spreadsheet users but also novice users. This more extensive survey will allow for a more general understanding of the need for mobile spreadsheet apps.

It is also planned to extend the current survey to examine a more extensive range of spreadsheet features and functions and to determine which of these are most important for mobile spreadsheet users. With this understanding it would be possible to design a mobile spreadsheet app that is optimised for the real needs of users and therefore puts people at the centre of mobile application development.

To better understand the issues that are experienced by mobile spreadsheet users, it is intended to run a controlled experiment which asks participants to use a mobile spreadsheet app to retrieve information from an existing spreadsheet and to create a new spreadsheet on the device. By recording the users' actions while they are performing these tasks it will be possible to see what procedures users employ while interacting with the device and to see what errors they make during these common spreadsheet tasks.

Using this information a deeper insight into user behaviour can be gained and used to make recommendations as to how mobile spreadsheet apps should be designed to produce a more usable mobile spreadsheet app.

## 8 CONCLUSIONS

This research has shown that there is a need for mobile spreadsheet apps. A large proportion of participants surveyed, 78%, have said that they have needed access to spreadsheets while away from traditional computing devices. The primary needs of these users are to view or change existing spreadsheets with only 21% of participants saying they would need to create a spreadsheet on a mobile device.

Along with examining the need for mobile spreadsheet applications, this work also examined the extent to which existing applications are used. It was found that approximately half of the participants that required access to spreadsheets while away from traditional computing devices have used mobile spreadsheet applications. Of these participants only 33% could accomplish their desired task, with a further 50% stating that they could partially complete their task. To better understand why this was the case, the usability of these applications was examined.

It was found that existing apps suffer from a number of issues, most predominantly due to the small screen size of most mobile devices. Participants' most common complaint of these devices was that the screen size was too small. One participant also remarked that a lack of a global view was a problem. Despite these issues, it was found that participants were not dissatisfied with mobile applications. The convenience afforded by the ability to view spreadsheets on a mobile device outweighs the usability issues associated with mobile spreadsheet applications.